\documentclass[sigconf]{acmart}

\usepackage[applemac]{inputenc}
\usepackage[T1]{fontenc} 
\usepackage[english]{babel}
\usepackage{amsmath}
\usepackage{graphicx,tikz}
\usepackage{array,multirow}
\usepackage{caption,subcaption}

\theoremstyle{definition}
\newtheorem{definition}{Definition}

\newtheorem{proposition}[definition]{Proposition}
\newtheorem{definition/proposition}[definition]{Definition/Proposition}
\newtheorem{theorem}[definition]{Theorem}

\newtheorem{example}[definition]{Example}

\title{Group Fairness: Independence Revisited}

\author{Tim Räz}

\affiliation{\department{Institute of Philosophy} \institution{University of Bern}\country{Switzerland}}

\affiliation{\department{Institute of Biomedical Ethics and History of Medicine} \institution{University of Zürich}\country{Switzerland}}

\email{tim.raez@posteo.de}

\copyrightyear{2021} 
\acmYear{2021} 
\setcopyright{acmlicensed}\acmConference[FAccT '21]{ACM Conference on Fairness, Accountability, and Transparency}{March 1--10, 2021}{Virtual Event, Canada}
\acmBooktitle{ACM Conference on Fairness, Accountability, and Transparency (FAccT '21), March 1--10, 2021, Virtual Event, Canada}
\acmPrice{15.00}
\acmDOI{10.1145/3442188.3445876}
\acmISBN{978-1-4503-8309-7/21/03}

\keywords{fairness, independence, statistical parity, demographic parity, sufficiency, separation, affirmative action, accuracy}

\ccsdesc[500]{Social and professional topics~User characteristics}
\ccsdesc[300]{Computing methodologies~Machine learning}
\ccsdesc[300]{Applied computing~Arts and humanities}

\begin{document}

\begin{abstract}
This paper critically examines arguments against \emph{independence}, a measure of group fairness also known as \emph{statistical parity} and as \emph{demographic parity}. In recent discussions of fairness in computer science, some have maintained that independence is not a suitable measure of group fairness. This position is at least partially based on two influential papers (Dwork et al., 2012, Hardt et al., 2016) that provide arguments against independence. We revisit these arguments, and we find that the case against independence is rather weak. We also give arguments in favor of independence, showing that it plays a distinctive role in considerations of fairness. Finally, we discuss how to balance different fairness considerations.
\end{abstract}

\maketitle

\section{Introduction}

Measures of group fairness have become an important topic in computer science after the publication of the ProPublica article ``Machine Bias'' \cite{angwi2016}. ProPublica found that the risk assessment tool COMPAS is biased against black people in having unbalanced false positive and false negative rates. This is intuitively unfair. The ensuing debate mostly focused on the contrast between the measure implicitly used by ProPublica, now known as \emph{separation}, and other measures, in particular a measure known as \emph{sufficiency}. However, a third measure of group fairness, \emph{independence}, also known as \emph{statistical parity} or \emph{demographic parity}, has been viewed more critically. Some computer scientists seem to think that independence is not a suitable measure of group fairness \citep{klein2016,berk2018}; others maintain that while independence is adequate in some contexts, it leads to undesirable consequences in others \citep{zemel2013,choul2017}. The critical stance of computer scientist with respect to independence appears to be at least partially based on two influential papers \citep{dwork2012,hardt2016} that provide arguments against independence. Here we revisit and critically examine these arguments, and we find that the case against independence as opposed to other measures of group fairness is rather weak.

We first introduce measures of group fairness and their most important properties (section \ref{sec: fairness criteria def.}). In particular, we introduce the concept of \emph{conservative} fairness measures, which allows us to clarify the relation between fairness and accuracy. We then examine the arguments against independence (section \ref{sec: against independence}). We find that, first, arguments against independence proposed in \cite{dwork2012} equally apply to other measures of group fairness such as sufficiency and separation, and should therefore not be taken to apply to independence specifically. Second, we argue that arguments against independence proposed in \cite{hardt2016} are flawed in making unwarranted assumptions about conservative fairness measures such as sufficiency and separation. We prove that sufficiency and separation are not \emph{incrementally conservative}, which means that these measures are not necessarily preserved if we increase the accuracy of a predictor. We then state arguments in favor of independence (section \ref{sec: pro independence}), finding that independence captures aspects of fairness not covered by sufficiency and separation. Finally, we discuss how to balance different fairness considerations (section \ref{sec: justifying tradeoffs}).

\section{Fairness Measures: Definitions and Properties}
\label{sec: fairness criteria def.}

This section introduces and discusses the most important measures of group fairness, formulates these measures for the case of binary variables, and discusses other relevant fairness measures, setting the stage for the discussion in later sections.

\subsection{Definitions of Group Fairness}

Here we state the most important group fairness measures, following the discussion in \citep{baroc2019}. These measures are formulated using random variables $Y$, $R$, $A$; all measures we consider correspond to statistical properties of these variables. The variables have the following interpretation: $Y$ is the ``true label'', i.e., the characteristic that we want to predict; $R$ is the prediction, which can be the output of an algorithm; $A$ is the characteristic indicating group membership, i.e., the property with respect to which we investigate fairness. In the context of supervised learning, we have access to $Y$ through labeled data. We will mostly focus on binary variables. Also, we will assume that a prediction $R$ leads to a corresponding decision.

Let us illustrate this setup using the example of college admissions. College students of different genders apply for college; a prediction about their suitability is made based on the application documents. In this case, the value of $Y$ corresponds to the actual suitability of a student applying for college. $Y$ is known if the data in question is historical, and the value of $Y$ can be determined based on whether a student actually obtained a degree or not (or a different operationalization of `suitable applicant'). $R$ corresponds to the prediction whether or not a student should be admitted to college based on the application documents. $A$ corresponds to the gender of applicants, which we assume to be binary for simplicity's sake.

To formulate fairness measures, we will use the following notation: Two random variables $X,Y$ are independent if $P(X,Y) = P(X)\cdot P(Y)$; we will write this as $X \perp Y$. Two random variables $X, Y$ are conditionally independent given $Z$ if $P(X\mid Y, Z) = P(X\mid Z)$; we will write this as $X\perp Y \mid Z$.

\begin{definition}
The measure of \emph{indepencende} is satisfied if $R\perp A$.
\label{independence}
\end{definition}

Independence, also known as statistical parity and demographic parity, means that the prediction $R$ does not depend on $A$. If independence is satisfied, a prediction is statistically balanced between different groups, in that members of the different groups get predictions at the same rate. In the case of college admissions, this means that an equal proportion of men and women applying for college are predicted to be suitable applicants.

\begin{definition}
The measure of \emph{sufficiency} is satisfied if $Y \perp A \mid R$.
\end{definition}

Sufficiency means that, given the prediction, the true label is independent of the group. The idea is that the prediction $R$ contains all the information about the true label, so the sensitive characteristic is not needed; in other words, the prediction $R$ is sufficient for $Y$. In the case of college admissions, this means that an equal proportion of men and women predicted to be suitable applicants are actually suitable applicants.\footnote{Sufficiency is closely related to \emph{calibration}. Calibration means that the predicted store reflects the true score. Calibration and sufficiency are equivalent up to reparametrization, cf. \citep[p. 52]{baroc2019}.}

\begin{definition}
The measure of \emph{separation} is satisfied if $R \perp A \mid Y$.
\end{definition}

Separation means that, given the true label, the prediction is independent of the group. The idea is that the prediction $R$ can only vary with respect to different groups $A$ insofar as this is justified by the true label $Y$; see \citep{baroc2019}. In the case of college admissions, this means that an equal proportion of suitable men and women applying for admission are predicted to be suitable applicants.

\subsection{Properties and Relations}
\label{sec: properties and relations}

In this section, we discuss some important properties of the fairness measures introduced above. The discussion follows \citep{baroc2019}; see the appendix for proofs. First, the \emph{accuracy} of a predictor $R$ is the degree to which it agrees with the true label $Y$; a perfect predictor is a predictor that completely agrees with the true label, i.e., $Y = R$. Next, we state an important property that is shared by sufficiency and separation, but not by independence.

\begin{proposition}
If we have a perfect predictor, then sufficiency and separation hold.
\label{ref: perfect pred}
\end{proposition}

Independence, $R \perp A$, is not, in general, compatible with a perfect predictor: independence and perfect predictors are only compatible if the true label is evenly distributed between groups, i.e., if we have $Y \perp A$, which is not the case in general. Proposition \ref{ref: perfect pred} motivates a definition that will be important in the following. It is a distinction between different kinds of group fairness measures:

\begin{definition}
A fairness measure is \emph{conservative} if the measure is necessarily satisfied in the case of a perfect predictor. Otherwise, a fairness criterion is \emph{non-conservative}.
\end{definition}

Fairness measures are called conservative because, in the case of a perfect predictor, they do not force us to change anything to obtain fairness, i.e., they conserve the \emph{status quo}.\footnote{The notion of a conservative fairness measure used here is related to the concept of \emph{conservative justice} \cite[Sec. 2.1.]{mille2017} insofar as the latter notion concerns the preservation of (factual) practices; however, the notion of conservativeness proposed here does not concern the preservation of norms as required by conservative justice.} Proposition \ref{ref: perfect pred} shows that both sufficiency and separation are conservative fairness criteria; meanwhile, independence is not. Note that the reverse implication of proposition \ref{ref: perfect pred} is false. The following proposition provides a characterization of when sufficiency and separation holds in some cases of non-perfect predictors:

\begin{proposition}
If the joint distribution of $(A,Y,R)$ is positive for all values, then sufficiency and separation hold at the same time iff. $A$ is independent of the joint distribution of $Y$ and $R$, i.e., if $A \perp (Y, R)$.
\label{prop: non-perfect sep suff}
\end{proposition}

This proposition is important because it tells us when sufficiency and separation hold under reasonable circumstances such as a non-vanishing joint distribution.

The notion of a conservative fairness measure is very strong and of limited practical relevance, because predictors are hardly ever perfect in practice. To overcome this limitation, we can define a broader notion of conservativeness and investigate if relevant fairness measures are conservative on this notion:

\begin{definition}
A fairness measure is \emph{incrementally conservative} if the degree to which the measure is satisfied does not decrease if we increase the accuracy of the predictor.
\end{definition}

Are the fairness measures we considered above conservative according to this broader notion? Unfortunately, this is not the case. In the appendix, the following proposition is proved:

\begin{proposition}
Sufficiency and separation are not incrementally conservative fairness measures.
\label{prop: incremental conservative}
\end{proposition}

This means that if these two measures are satisfied by a certain (non-perfect) predictor, and we increase the accuracy of that predictor, it can happen that the improved predictor no longer satisfies the two measures. Thus, the property of conservativeness does not imply incremental conservativeness; it is not necessarily the case that if we increase the accuracy of a predictor, a conservative fairness measure is preserved.

\subsection{Confusion Matrices}
\label{sec: confusion matrices}

In this section, we discuss group fairness measures in the case of binary prediction $R$, ground truth $Y$, and characteristic $A$, using so-called confusion matrices. Confusion matrices make it easier to formulate and reason about these measures in concrete applications. The discussion in this section draws on more thorough expositions of confusion matrices and their characteristics in \cite{berk2018,loi2019}.

Assume we have collected statistical information for binary $Y$ and $R$, for example, historical records of college success $Y$, as well as the binary prediction for admission $R$. The prediction $R$ can be positive or negative, and for either outcome, it can match the true label $Y$ ($a$ = \# of true positives, $d$ = \# of true negatives) or be mistaken ($b$ = \# of false positives, $c$ = \# of false negatives). Based on these records, we can compile a confusion matrix:

\begin{figure}[h]
\centering
\begin{tabular}{l|l|c|c|c}
\multicolumn{2}{c}{}&\multicolumn{2}{c}{true label (Y)}&\\
\cline{3-4}
\multicolumn{2}{c|}{}&positive&negative&\multicolumn{1}{c}{total}\\
\cline{2-4}
\multirow{2}{*}{prediction (R)}& positive & $a$ & $b$ & $a+b$\\
\cline{2-4}
& negative & $c$ & $d$ & $c+d$\\
\cline{2-4}
\multicolumn{1}{c}{} & \multicolumn{1}{c}{total} & \multicolumn{1}{c}{$a+c$} & \multicolumn{    1}{c}{$b+d$} & \multicolumn{1}{c}{$N$}\\
\end{tabular}
\end{figure}

On this basis, we can define some important statistics of confusion matrices:

\begin{itemize}

\item Accuracy: $\frac{a + d}{N}$

\item Positive Predictive Value (PPV): $\frac{a}{a + b}$ 

\item Negative Predictive Value (NPV): $\frac{d}{c + d}$

\item False Positive Rate (FPR): $\frac{b}{b + d}$

\item False Negative Rate (FNR): $\frac{c}{a + c}$

\end{itemize}

From here on, we assume that we have observed sufficiently many cases such that the observations (relative frequencies) in our tables approximately match the ``true probabilities''. Now, in order to formulate fairness measures for confusion matrices, we need one matrix for each of two groups $p, q$ (values of the random variable $A$):

\vspace{0.2cm}

\begin{minipage}{1.5in}
\center
\begin{tabular}{l|c|c|c|c}
\multicolumn{2}{c}{}&\multicolumn{2}{c}{truth (Y)}&\\
\cline{3-4}
\multicolumn{2}{c|}{}&+&--&\multicolumn{1}{c}{}\\
\cline{2-4}
\multirow{2}{*}{pred. (R)}&+& $a$ & $b$ & $$\\
\cline{2-4}
&--& $c$ & $d$ & $$\\
\cline{2-4}
\end{tabular}
\vspace{0.1cm}
\captionof{table}{Group A=p}
\label{table: A}
\end{minipage}
\begin{minipage}{1.5in}
\center
\begin{tabular}{l|c|c|c|c}
\multicolumn{2}{c}{}&\multicolumn{2}{c}{truth (Y)}&\\
\cline{3-4}
\multicolumn{2}{c|}{}&+&--&\multicolumn{1}{c}{}\\
\cline{2-4}
\multirow{2}{*}{pred. (R)}&+& $a'$ & $b'$ & $$\\
\cline{2-4}
&--& $c'$ & $d'$ & $$\\
\cline{2-4}
\end{tabular}
\vspace{0.1cm}
\captionof{table}{Group A=q}
\label{table: A'}
\end{minipage}

Based on this, the three group fairness measures defined in the previous section can be formulated in terms of statistics of these two confusion matrices:

\begin{proposition}
For binary variables $Y, R, A,$ independence is equivalent to: $\frac{a+b}{N} = \frac{a'+b'}{N'}$.
\end{proposition}

\begin{proposition}
For binary variables $Y, R, A,$ sufficiency holds iff. both groups have the same positive predictive value (PPV), i.e., $\frac{a}{a + b} = \frac{a'}{a' + b'}$ and the same negative predictive value (NPV), i.e., $\frac{d}{c + d} = \frac{d'}{c' + d'}$.
\label{prop: sufficiency}
\end{proposition}

\begin{proposition}
For binary variables $Y, R, A,$ separation holds iff. both groups have the same false positive rate (FPR), i.e., $\frac{b}{b + d} = \frac{b'}{b' + d'}$ and false negative rate (FNR), ie, $\frac{c}{a + c} = \frac{c'}{a' + c'}$.
\label{prop: separation}
\end{proposition}

\subsection{Other Kinds of Fairness}
\label{sec: fairness other}

In this section, we discuss kinds of fairness that are not group fairness measures, but that will play an important role in our discussion below. The group fairness measures introduced above are observational measures, i.e., they can be measured based on data that are typically available: In may cases, we have access to labeled data ($Y$), a predictive model ($R$), and labels or a different kind of access to the sensitive characteristic of individuals, ($A$). However, there are other kinds of fairness considerations that are not measurable in terms of these quantities. 

Kamishima et al. \citep{kamis2011} propose to distinguish three different kinds of fairness. The first kind, prejudice, subsumes the notions of group fairness discussed above. The second kind, underestimation, is due to the fact that a model may be unfair due to the finiteness of training data. The definition of the third kind, negative legacy, is particularly important:

\begin{definition}
\emph{Negative legacy} is unfairness due to unfair sampling or labeling in the training data.
\end{definition}

Kamishima et al. provide the following example of negative legacy: ``[I]f a bank has been unfairly rejecting the loans of the people who should have been approved, the labels in the training data would become unfair. This problem is serious because it is hard to detect and correct'' (Ibid., p. 646). Kamishima et al. note that the problem can be overcome to a certain extent if an independent set of fairly labeled training data is available.

A further notion of fairness that is relevant is individual fairness, which can be defined as follows:

\begin{definition}
(Informal) \emph{Individual fairness} is the requirement that a fair predictor should treat similar individuals similarly, i.e., their predictions should be similar.\footnote{The notion is due to \citep{dwork2012}. Formally, we can make individual fairness precise by replacing the informal notion of ``similarity'' with two metrics, which capture how similar or close individuals and their predictions are. To do this, we need a metric $d$ between individuals $x, y \in I$, and a metric $D$ between distributions of predictions $Mx, My$ of individuals, where $M$ is a map from individuals $I$ to distributions of predictions. To enforce individual fairness, we now require that the distance between individuals should limit the distance between the distribution of predictions, i.e., we should have $D(Mx,My) \leq d(x,y)$ for $x, y \in I$. This is a so-called Lipschitz condition.}
\end{definition}

Individual fairness is in tension with group fairness measures under certain conditions because group fairness defines fairness in terms of (average) properties of group members, which usually does not do justice to some individual properties of group members. In particular, if the comparison of individuals uses fine-grained information such as a score or utility, it is possible to violate individual fairness while complying with some measure of group fairness. We will see examples of this below. Finally, note that individual fairness seem to be the fairness measure that is most closely related to the philosophical concept of justice, cf. \cite[Sec. 1.1.]{mille2017}

\section{Arguments Against Independence Scrutinized}
\label{sec: against independence}

In this section, we examine arguments against independence from the computer science literature. When following the debate, one can get the impression that independence is somehow flawed or unsuitable as a measure for group fairness. The goal of this section is to revisit and critically examine important arguments against independence.

\subsection{Dwork et al.: The Argument From Gerrymandering}

The most important paper credited with showing that independence is not a suitable fairness concept is \citep{dwork2012}. We will reexamine the arguments in this paper and argue that, first, it is not clear whether Dwork et al. wish to reject independence, and second, that the arguments made by Dwork et al. should not be construed as arguments against independence, but more broadly as arguments against group fairness in general.

First, let us examine whether Dwork et al. wish to simply reject independence. Note that Dwork et al. call independence ``statistical parity''. In the introduction, Dwork et al. write: ``we demonstrate [the inadequacy of statistical parity] as a notion of fairness through several examples in which statistical parity is maintained, but from the point of view of an individual, the outcome is blatantly unfair.'' (p. 2) This sounds like an outright rejection. However, later in the paper, Dwork et al. write that ``statistical parity is insufficient as a general notion of fairness.'' (p. 7) This suggest that Dwork et al. merely want to argue that independence (statistical parity) is not a logically sufficient condition for fairness, which is a much weaker claim than the claim that it should be rejected \emph{tout court}. What is more, the paper investigates to what extent individual fairness implies, or helps satisfy, independence, which would be unnecessary if independence should be outright rejected. Thus, Dwork et al. merely caution against independence as the sole arbiter of fairness.

Now let us examine the arguments against independence in \citep[Sec. 3.1]{dwork2012} more closely. The arguments take the form of three examples, in which the adoption of independence has undesirable consequences, in that independence holds, but individuals are treated unfairly, that is, individual fairness is violated. The first example, `Reduced Utility', shows that independence does not ensure that the most suitable candidates from different group are selected. In the example, an organization hires people from two groups $p, q \in A$. It is possible for the organization to comply with independence while, out of ignorance, choosing the least qualified members of group $p$ and the best qualified members of group $q$. This reduces the utility of the organization, and it also violates individual fairness, because similar members of the two groups are treated differently. To make it concrete, assume we have two individuals $x \in p$ and $y \in q$, both similarly qualified, and while $y$ is hired, $x$ is not hired; thus two individuals who are similar are not treated similarly. The second example, `Self-fulfilling Prophecy', has the same structure as the first example, but the unqualified members of $p$ are now maliciously chosen for the purpose of justifying future discrimination of members of $p$. The third example, `Subset Targeting', is based on the fact that independence does not ensure a fair choice within groups, in that it does not require that the most deserving members of a group get to see a relevant job ad. This implies, once more, a violation of individual fairness.

Are these examples by Dwork et al. sufficient to reject independence as a criterion of group fairness? We grant that independence can decrease utility, if utility depends on the degree of accuracy, because independence does not depend on $Y$, and thus also not on accuracy, i.e., how well $Y$ and $R$ match. The three examples discussed by Dwork et al. are all based on the fact that independence only requires that an equal proportion of two groups get classified in a certain way, but does not further specify how individuals within these groups have to be distributed with respect to $Y$. As a consequence, independence does not, in general, guarantee individual fairness.

We thus grant that these examples are valid in substance. However, this is not sufficient to reject independence as opposed to other measures of group fairness such as separation and sufficiency, because similar arguments can be directed against these other measures. Our argumentative strategy is to make a \emph{tu quoque} argument: If one accepts these arguments against independence, then one also has to accept similar arguments against sufficiency and separation.

We will now give a first example of gerrymandering \cite{kearn2018} with separation, in which an employer manipulates statistics so as to realize an unequal treatment of groups, while maintaining separation.

\begin{example}
\emph{Gerrymandering With Separation:} A malicious employer makes hiring decisions. There are two groups, $p$ and $q$, for which separation has to be enforced. The employer has a preference for people from group $q$. Assume that the employer has made provisional hiring decisions that satisfy separation, and a confusion matrix according to these hiring decisions has been compiled, cf. tables \ref{table: A} and \ref{table: A'}. The confusion matrices satisfy separation, which implies that we have $\frac{a}{a+c} = \frac{a'}{a'+c'}$. Assume further that the employer has a reservoir of $z$ qualified candidates from group $q$ that do not appear in the statistic of the confusion matrix. The employer can now hire candidates from that reservoir, as long as an appropriate proportion of qualified candidates is rejected, i.e., the employer creates a division $z = z_+ + z_-$ into qualified people hired $z_+$ and qualified people rejected $z_-$, such that $\frac{a' + z_+}{a'+c' + z} = \frac{a'}{a'+c'}$. The new confusion matrices are unchanged except for the entries $a' \rightarrow a' + z_+$ and $c' \rightarrow c' + z_-$, which means that the matrices still satisfy separation, as can be easily verified. However, this hiring practice seems to be intuitively unfair towards group $p$ (and the qualified people from the reservoir $z$ who are rejected); it can also violate individual fairness because equally suitable candidates from group $p$ are not even considered.
\end{example}

It could be objected that this example is not analogous to the examples by Dwork et al. in that here, the employer has to keep part of the statistic ``off the book'', and add it later. However, the examples by Dwork et al. also make the assumption that there is additional information not captured by the variables relevant to independence. It is important to note that this (malicious) hiring practice would not be possible in the case of independence, because in the example, the employer drives up the number of employees from group $q$, without raising the numbers in the other group, and independence enforces exact balance between groups. 

Let us give a second example, with a different structure. This is an example for gerrymandering with sufficiency and separation.

\begin{example}
\emph{Gerrymandering With Sufficiency and Separation:} Assume that an employer has made provisional hiring decisions and compiled two confusion matrices. Assume that the confusion matrices for $p$ and $q$ both do not have any zero entries and that the two confusion matrices have the same entries if they are normalized by $N$ and $N'$ respectively; this means that the corresponding joint distribution of $(Y,R,A)$ is positive everywhere, and the joint distribution of $(Y,R)$ is independent of group membership $A$. In this case, sufficiency and separation are both satisfied according to proposition \ref{prop: non-perfect sep suff}. Now, the employer wants to hurt group $p$. Under certain circumstances, this can be done as follows. The employer chooses a member $x$ of $p$ that is a false negative, i.e., $x$ should be hired, but is not predicted to be hired. If there is a member $x^*$ in group $p$ that is a true positive, i.e., should be hired, and is predicted to be hired, and is more suitable for the job than $x$, the malicious employer can simply switch the prediction for $x$ and $x^*$, such that $x$ becomes a true positive and $x^*$ becomes a false negative. The two confusion matrices are unchanged and thus still satisfy sufficiency and separation. However, individual fairness is violated, because a less qualified candidate has been chosen over a more qualified candidate, which hurts the group.
\end{example}

Note that this example also works without assuming (malicious) intent. The situation described arises naturally if an employer has less information about the relative suitability of people from group $p$; the employer would then hire a less than optimal selection of candidates from group $p$ and thus also not maximize utility. This is very similar to the first example by Dwork et al.. We can conclude that examples of gerrymandering that are similar to those by Dwork et al., can be constructed for notions of group fairness such as separation and sufficiency. Note that other examples of gerrymandering for sufficiency and separation can be constructed along the lines of the examples given here.

It could be asked why Dwork et al. did not appreciate that their arguments apply to other notions of group fairness as well. Here is a plausible explanation: Except for independence, notions of group fairness, such as sufficiency and separation, were only discussed more widely after the publication of the seminal ProPublica article \citep{angwi2016}, which appeared four years after the publication of \citep{dwork2012}. Thus, Dwork et al. might have raised their objections against notions of group fairness in general, and not just against independence, if they had been aware of other notions. It should also be noted that the primary focus of Dwork et al. is on the notion of individual fairness, not on group fairness. The paper does examine the relation between individual fairness and independence (statistical parity), but this is not the center of attention. All in all, the idea that independence is somehow more problematic than other group fairness criteria, which is held in parts of the computer science literature and can at least partially be traced back to Dwork et al., may be a historical accident.\footnote{\citep{baroc2019} mention that argument against independence may apply to other statistical fairness measures as well, but this is not elaborated.}

\subsection{Hardt et al.: The Argument From Accuracy}

Other arguments in the computer science literature are targeted more specifically at independence and do not apply to other notions of group fairness such as sufficiency and separation. In \citep{hardt2016}, the authors propose separation as a fairness measure. To differentiate separation from independence, the authors claim that independence is flawed for two reasons that do not apply to separation. The first reason is the kind of argument made in \citep{dwork2012}. The second reason why independence is flawed is given in the following quote -- note that the authors call independence ``demographic parity'', and the predictor is denoted $\hat{Y}$:

\begin{quotation}
... demographic parity often cripples the utility that we might hope to achieve. Just imagine the common scenario in which the target variable $Y$ -- whether an individual actually defaults or not -- is correlated with $A$. Demographic parity would not allow the ideal predictor $\hat{Y} = Y$ , which can hardly be considered discriminatory as it represents the actual outcome. As a result, the loss in utility of introducing demographic parity can be substantial. \citep[p. 2]{hardt2016}
\end{quotation}

Later, the authors note that separation does not have this problem: ``Unlike demographic parity, our notion always allows for the perfectly accurate solution [...]'' (Ibid.) We will now reconstruct the argument in this passage based on the distinctions made in section \ref{sec: fairness criteria def.}. There are two different readings of the argument. The first reading focuses on the relation between accuracy and utility, while the second reading focuses on the relation between accuracy and fairness. On the first reading, the argument can be reconstructed as follows:

\begin{itemize}

\item[P1] A perfect predictor maximizes utility.

\item[P2] Independence is a non-conservative fairness criterion (is not generally compatible with a perfect predictor), while separation is a conservative fairness criterion (is compatible with a perfect predictor).

\item[C1] Therefore, independence is not generally compatible with maximal utility, while separation is.

\item[C2] Therefore, separation should be preferred over independence.

\end{itemize}

There are two main problems with this argument. The first problem is premiss 1: It is not the case that accuracy and utility align necessarily; see \cite{corbe2017}. For one, accuracy only captures the state of the world as it is at a certain point in time. Thus, if we maximize accuracy, we maximize utility only with regard to short-term goals. To take the example of risk assessment, maximizing utility means minimizing current risk. This does not take into account the value of changing risk assessment so as to minimize, say, future risk, which can be tied to, say, racial justice. It is explicitly noted in \cite{corbe2017} that utility according to present risk scores is ``immediate utility''. Furthermore, note that if we have a predictor $R$ that is not perfect, and false positives and false negatives have different utilities, we may have to choose a predictor $R'$ that is even less accurate than $R$ to maximize utility.

The second problem is the step from conclusion 1 to conclusion 2. As is often pointed out in the computer science literature, we virtually never have a perfect predictor. So we are almost never in a situation where it actually matters that a fairness measure is conservative, i.e., that the measure is compatible with the perfect predictor. However, if we are almost never in this situation, conservativeness is a theoretical concern, but practically irrelevant. A situation that is practically irrelevant should not guide our choice of fairness measure. So, there is no practical reason to prefer conservative fairness measures over non-conservative ones.\footnote{Note that from a conceptual or philosophical point of view, it could be worthwhile to explore the case of perfect predictors. The argument made here takes the more practical position of computer science that perfect predictors are negligible as a point of departure.}

It could be thought that the above argument also goes through for broader notions of conservativeness, i.e., that it holds for increments of accuracy: if we increase accuracy, and this automatically increases the degree to which a fairness measure holds, then we do not need a perfect predictor for accuracy to be of practical relevance; the two align in increments. In fact, Hardt et al. appear to have an argument along these lines in mind. Immediately after the passage quoted above, they write: 

\begin{quotation}
[O]ur criterion is easier to achieve the more accurate the predictor $\hat{Y}$ is, aligning fairness with the central goal in supervised learning of building more accurate predictors. \cite[p. 2]{hardt2016}
\end{quotation}

This claim, however, is false in view of proposition \ref{prop: incremental conservative}, which establishes that it is possible to start with a predictor $R$ that satisfies separation, increase the accuracy of $R$, and obtain a new predictor $R'$ that no longer satisfies separation. Proposition \ref{prop: incremental conservative} shows that both separation and sufficiency are not incrementally conservative, and that, therefore, an incremental version of the above argument does not support separation or sufficiency as opposed to independence.

Let us now turn to the second reading of the argument in the quote from Hardt et al., which focuses on the relation between accuracy and fairness: 

\begin{itemize}

\item[$P1^*$] A perfect predictor is (maximally) fair, because it aligns with the actual outcome.

\item[$P2^*$] Independence is a non-conservative fairness criterion (is not generally compatible with a perfect predictor), while separation is a conservative fairness criterion (is compatible with a perfect predictor).

\item[C1*] Therefore, independence is not generally compatible with a (maximally) fair predictor, while separation is.

\item[C2*] Therefore, separation should be preferred over independence.

\end{itemize}

There are, again, two problems with this argument. The first problem, the step from the first to the second conclusion, was already discussed above -- we can reasonably doubt the practical relevance of perfect predictors, because they are virtually never realized, and an incremental version of the argument is demonstrably false. The second, more fundamental problem is premiss $P1^*$. This premiss is unsupported, and, arguably, wrong in general. Premiss $P1^*$ is problematic both from a philosophical and from a computer science perspective.

From a computer science perspective, there are important aspects of algorithmic fairness that are not captured by group fairness measures, and this is well known. Take, for example, the kinds of fairness discussed in \citep{kamis2011}; see section \ref{sec: fairness other} above. \emph{Negative legacy} is unfairness due to unfair sampling or labeling. Consider the case of unfair labeling. Unfair labeling means that the distribution $P(Y,A)$ is unfair, i.e., the distribution of actual outcomes $Y$ we measure at a certain point in time favors one of the groups in $A$ over another in a way we consider to be unfair. What premiss $P1^*$ says is that a perfect predictor $R$ is fair because it aligns with the actual outcome, i.e., because we have $Y = R$. However, $Y = R$ only provides a good justification of the fairness of $R$ with respect to $A$, i.e., of $P(R,A)$, if the distribution $P(Y,A)$ itself is fair, which need not be the case if labeling is unfair; this is what Kamishima et al. point out. The distribution $P(Y,A)$ can arise through unfair practices, historical biases, and so on. 

Importantly, Kamishima et al. also point out that this sort of unfairness is hard to detect or measure if we do not have access to a sample with fair labeling, such that we can obtain a fair estimate of $P(Y,A)$. But of course, just because it can be hard, or even impossible, to quantify negative legacy, does not mean that this quantity is of no ethical import. Fairness is completely independent of our ability to measure it.

Let us illustrate these points with some examples. Why should we think that an accurate predictor is fair? One of the reasons may be that an accurate predictor aligns with the ground truth $Y$. And trying to align predictions with the truth should not be considered to be discriminatory -- this is the point made by Hardt et al. in the above quote. To address this point, recall what truth means in the present context: It means that $Y$ captures what we observe in the world at a certain point. For example, we observe that people from group $p$ in fact get arrested more frequently than people from group $q$, we observe that group $p$ in fact has more loan applications rejected than group $q$, and so on. This is what the joint distribution of $Y$ and $A$ captures. In other words, the distribution $P(Y,A)$ is a picture of the status quo. However, the world as it is at a certain point, or the status quo, is not a moral category. It is just a description of what we find in the world. It does not answer the question whether the world as we find it is fair, or morally justified. Finding the world to be a certain way, and inferring from this that the world ought to be this way, is committing a fallacy according to some philosophers, based on a confusion between facts and values; see, e.g., the discussion of the Is-Ought gap in \cite[Sec. 2.1.]{vaery2019}.

At this point, it could be objected that in some cases, the distribution of labels does have moral import. Take, for example, the often-mentioned case of violent offenders. If the distribution $P(Y,A)$ captures the historical record of reoffending of violent criminals in the past, then it makes sense to align our predictor $R$ with $Y$. It seems that we cannot just ignore the historical record in favor of a group fairness measure such as independence. The price we pay by releasing (potentially) violent criminals from one group, or by locking up (potentially) innocent members of the other group because these groups have different frequencies with respect to $Y$, seems very high, and the choice of such a predictor seems morally wrong. A form of this argument is made in the following passage of  \citep[p. 14]{berk2018}: ``[Independence] has been criticized because it can lead to highly undesirable decisions for individuals (Dwork et al. 2012). One might incarcerate Muslims who pose no public safety risk so that the same proportions of Muslims and Christians are released on parole.''

The response to this objection is that it is perfectly possible that ignoring the status quo has undesirable moral \emph{consequences}, as in the case of violent offenders. However, this does not invalidate the point that the status quo \emph{in itself} does not have moral status. It just means that the status quo can impact considerations of fairness in some cases, and that we may have to weight the moral consequences of sticking to or deviating from the status quo against other considerations of fairness. We will turn to a discussion of how this could be achieved in the next section.

\section{Arguments in Favor of Independence}
\label{sec: pro independence}

So far, we have examined arguments against independence, and we have found that the case against independence is not as clear cut as some of the computer science literature suggests. In this section, we turn to the case \emph{for} independence. Why is independence a good or useful fairness measure? We compare independence to other notions of group fairness to highlight its usefulness, but also its limitations. Our goal is not to recapitulate the philosophical literature that supports independence. Rather, our goal is to establish some connections between philosophical concerns and the more formal discussion in computer science.

Independence is defined as $R \perp A$, that is, probabilistic independence of group membership and prediction. Note that in practice, it makes sense to not require strict independence, but an approximate version of independence. One justification of independence is that it controls, and potentially compensates, for historical injustice. One manifestation of historical injustice is what Kamishima et al. call negative legacy \citep{kamis2011}, viz. a distribution $P(Y,A)$ that we consider to be unjust. The distribution can be unjust because it does not adequately represent the true properties of the groups involved -- this would correspond to unfair sampling, in which case we may not know the true distribution -- , or because the distribution does represent the true properties of the groups involved, but these properties themselves did not come about in a fair way -- this would correspond to unfair labeling.\footnote{Note that above, we have excluded the first case through the assumption that the confusion matrices are at least approximately representative of the true probabilities. We have not excluded the second case.}

Formally, negative legacy can manifest as a correlation between group membership $A$ and ground truth $Y$, i.e., $Y \not\perp A$: if the groups $A$ should have equal access to the outcome encoded by $Y$, there should be no correlation between group membership and outcome, i.e., we should have $Y \perp A$. Note that, as in the case of independence, we can formulate an approximate version of this requirement. Now, if we build a predictor $R$ with a focus on accuracy, as it is usually the case, we get $R \approx Y$, i.e., the predictor is approximately accurate. However, this also implies that the predictor $R$ does not satisfy (an approximate version of) independence. Thus, independence helps us detect this form of historical injustice, and it suggests that we modify $R$, such that, approximately, we obtain $R \perp A$. This modification of $R$ may also influence negative legacy in the long run by moving the distribution $Y$ closer to the desired $Y \perp A$ over time, such that accuracy and independence align naturally. This is one argument in favor of independence.

To better understand the usefulness of independence as a fairness measure, let us compare it to other kinds of measures. Take, first, sufficiency and separation. The main difference between independence on the one hand, and sufficiency and separation on the other, is that independence is formulated without $Y$. This means that while sufficiency and separation track the difference between a prediction $R$ and the truth given by $Y$ -- they are measures of error or deviation from the truth -- independence does not track deviation from the truth. \emph{Prima facie}, this may seem like a deficiency of independence. However, as was just explained, independence helps us detect unfairness in the distribution of $Y$ exactly because it does not focus on deviations from $Y$. It helps us to see what may be wrong with the distribution of $Y$ itself. This is an advantage of independence in contrast to separation and sufficiency.

Now let us compare independence to affirmative action, viz., the requirement that predictions $R$ have to satisfy certain thresholds or quota. In the case of college admissions, the requirement could be that a certain percentage of admitted candidates have to be members of a racial minority; see \cite{hertw2020} for a discussion of affirmative action in the context of college admissions. A justification for affirmative action is to compensate for historical injustice. In this respect, the justification of affirmative action is similar to the justification of independence given above. 

However, there are also important differences between independence and affirmative action. One difference is that independence only requires predictions to be independent of group membership. Affirmative action, on the other hand, can be more stringent in requiring that predictions satisfy certain proportions. For example, if only 10\% of college applicants belong to a minority, independence would require that the admission rate for these 10\% is the same as the general admission rate, while affirmative action may require that the admission rate among the 10\% is larger to allow for a given balance of admitted candidates, irrespective of application rates. This means that independence, formulated for a given set of applicants, will not correct for certain kinds of biases such as underrepresentation of groups among applicants, while affirmative action may correct for this kind of bias.

More generally, it should be stressed that while independence may highlight and help to compensate for certain kinds of historical injustice, implementing it will not correct for many other forms of injustice. In particular, independence prescribes an intervention only on the prediction $R$, which can be interpreted as a compensation for a certain distribution of $Y$, and does not prescribe an intervention on the causes of this distribution, or an intervention on the effects of this distribution.

\section{How to Reason About Fairness Measures}
\label{sec: justifying tradeoffs}

We have now seen arguments both in favor and against independence, and we have found that there is some validity to arguments on both sides. How should we proceed from here? How should these arguments be weighted? We will not be able to answer these questions here, but we can provide some rough guidelines in view of the above discussion.

First, we should always explicitly state the moral value of either choosing or rejecting a group fairness measure such as independence, as opposed to arguing solely on the basis of factual and descriptive properties of fairness measures. We have seen why this is important in the case of independence. We have argued that accuracy in and of itself does not have moral value. We do not deny that accuracy can be morally beneficial in certain situations or contexts; however, it is these moral benefits we care about, and they should be stated. For example, if neglecting accuracy has substantial social costs in some cases, this is what we care about, and not accuracy per se. Only once the values supporting arguments for or against independence have been made explicit can we weight them.

Second, gerrymandering is a problem shared by all measures of group fairness. It is possible to violate individual fairness while complying with sufficiency or separation, just as it is possible to comply with independence. Now, there is already a lot of work in computer science dealing with this problem, beginning with \cite{dwork2012}, who examine under which conditions independence and individual fairness can be combined. One of the problems of combining measures of group fairness and individual fairness will be, once more, to make the moral value of either choice explicit and assign appropriate weights to these choices.

Third, it is a mistake to think that we can either require or reject measures of group fairness independently of the case to which it is applied. Rather, the importance of different group fairness measures is context dependent. We have seen examples of this above: The cost of requiring independence in the case of classifying violent offenders is different from the cost of requiring independence in the case of college admissions. In the first case, the cost of making mistakes seems high; in the second case, the cost of making mistakes seems lower both for individuals and for society; see \citep{choul2017}.

Fourth, the preceding two points suggest that none of the group fairness measures we discussed here are logically necessary or logically sufficient for fairness: They cannot be logically sufficient because they violate individual fairness at least in some cases, and they cannot be logically necessary because they appear to be in conflict with our intuitions about fairness in other cases. This also suggests to interpret these measures of group fairness not as absolute criteria for fairness. Rather, they can be indicative of fairness or unfairness depending on the case at hand.

\section{Conclusion}

In this paper, we have examined the discussion of independence in the computer science literature, and we have found that some arguments against independence are not convincing in that they either equally apply to other measures of group fairness, or unduly emphasize descriptive properties of fairness measures, viz. conservativeness, as opposed to normative ones. We have also made a positive case for independence, arguing that it can highlight a distinct kind of unfairness not captured by sufficiency or separation. The main upshot of the present paper is that independence is an important measure of group fairness that has to be taken into account in discussions of algorithmic fairness.

\appendix

\section{Proofs}
\label{sec: appendix}

Here we give proofs of the propositions in the main text. All propositions and proofs can be found in the literature \citep{baroc2019,dawid1979,wasse2004} and are collected here for convenience's sake, except for the proof of proposition \ref{prop: incremental conservative}, which is new. We first state some useful properties of conditional independence (see the above references for proofs):

\begin{proposition}
Properties of conditional independence:

\begin{enumerate}

\item If $X \perp Y \mid Z$, then $Y \perp X \mid Z$;

\item if $X \perp Y \mid Z$ and $U = h(X)$, then i) $U \perp Y \mid Z$ and ii) $X \perp Y \mid (Z, U)$;

\item if $Y = h(Z)$, then $X \perp Y \mid Z$;

\item $X \perp Y \mid Z$ and $X \perp W \mid (Y,Z)$ iff. $X \perp (W,Y) \mid Z$;

\item if $X \perp Y \mid Z$, $X \perp Z \mid Y$, and (X,Y,Z) is positively distributed everywhere, then $X \perp (Y, Z)$.

\end{enumerate}

\end{proposition}

Note that properties 1, 2, 3 also hold without conditioning on $Z$.

\emph{Proof of proposition \ref{ref: perfect pred}:} A perfect predictor means that $Y = R$. Sufficiency means $A \perp Y \mid R$ and separation means $A \perp R \mid Y$. For a perfect predictor, these reduce to $A \perp Y \mid Y$. By property 3 of conditional independence, this is true for a perfect predictor because $Y = f(Y)$. $\square$ 

\emph{Proof of proposition \ref{prop: non-perfect sep suff}:} The direction (1) $\Rightarrow$ (2) is property 5 of conditional independence. The direction (2) $\Rightarrow$ (1) can be seen as follows: view $(Y,R)$ as a two-dimensional random variable, note that $Y$ and $R$ are functions of this random variable (projection), then the result follows from property 2 of conditional independence (without conditioning on $Z$). $\square$

\emph{Proof of Proposition \ref{prop: incremental conservative} (Incremental Conservativeness):}

We show that sufficiency and separation are not, in general, preserved if the accuracy of a predictor is increased, by giving an example where accuracy increases but separation and sufficiency are lost. First, consider the two following confusion matrices (recall that $Y$ stands for the true label, while $R$ stands for the prediction):

\vspace{0.2cm}

\begin{minipage}{1.5in}
\begin{tabular}{l|c|c|c|c}
\multicolumn{2}{c}{}&\multicolumn{2}{c}{Y}&\\
\cline{3-4}
\multicolumn{2}{c|}{}&+&--&\multicolumn{1}{c}{total}\\
\cline{2-4}
\multirow{2}{*}{R}&+& $10$ & $2$ & $12$\\
\cline{2-4}
&--& $3$ & $11$ & $14$\\
\cline{2-4}
\multicolumn{1}{c}{} & \multicolumn{1}{c}{total} & \multicolumn{1}{c}{$13$} & \multicolumn{ 1}{c}{$13$} & \multicolumn{1}{c}{}\\
\end{tabular}
\vspace{0.1cm}
\captionof{table}{Group A=p}
\end{minipage}
\begin{minipage}{1.5in}
\begin{tabular}{l|c|c|c|c}
\multicolumn{2}{c}{}&\multicolumn{2}{c}{Y}&\\
\cline{3-4}
\multicolumn{2}{c|}{}&+&--&\multicolumn{1}{c}{total}\\
\cline{2-4}
\multirow{2}{*}{R}&+& $20$ & $4$ & $24$\\
\cline{2-4}
&--& $6$ & $22$ & $28$\\
\cline{2-4}
\multicolumn{1}{c}{} & \multicolumn{1}{c}{total} & \multicolumn{1}{c}{$26$} & \multicolumn{1}{c}{$26$} & \multicolumn{1}{c}{}\\
\end{tabular}
\vspace{0.1cm}
\captionof{table}{Group A=q}
\end{minipage}

These matrices satisfy sufficiency and separation; the easiest way to see this is to check that the table for $q$ is a multiple of the table for $p$, so the relative frequencies are the same, which implies that sufficiency and separation are satisfied by proposition \ref{prop: non-perfect sep suff}. It can also be checked by hand, by using the relation between the statistics of confusion matrices on the one hand and fairness on the other, explained in section \ref{sec: confusion matrices}. Now we increase the accuracy of the predictor $R$, by taking, in each group, an element of the false negatives and shifting it to the true positives. This yields a new predictor $R'$ with the following confusion matrices:

\vspace{0.2cm}

\begin{minipage}{1.5in}
\begin{tabular}{l|c|c|c|c}
\multicolumn{2}{c}{}&\multicolumn{2}{c}{Y}&\\
\cline{3-4}
\multicolumn{2}{c|}{}&+&--&\multicolumn{1}{c}{total}\\
\cline{2-4}
\multirow{2}{*}{R'}&+& $11$ & $2$ & $13$\\
\cline{2-4}
&--& $2$ & $11$ & $13$\\
\cline{2-4}
\multicolumn{1}{c}{} & \multicolumn{1}{c}{total} & \multicolumn{1}{c}{$13$} & \multicolumn{ 1}{c}{$13$} & \multicolumn{1}{c}{}\\
\end{tabular}
\vspace{0.1cm}
\captionof{table}{Group A=p}
\end{minipage}
\begin{minipage}{1.5in}
\begin{tabular}{l|c|c|c|c}
\multicolumn{2}{c}{}&\multicolumn{2}{c}{Y}&\\
\cline{3-4}
\multicolumn{2}{c|}{}&+&--&\multicolumn{1}{c}{total}\\
\cline{2-4}
\multirow{2}{*}{R'}&+& $21$ & $4$ & $25$\\
\cline{2-4}
&--& $5$ & $22$ & $27$\\
\cline{2-4}
\multicolumn{1}{c}{} & \multicolumn{1}{c}{total} & \multicolumn{1}{c}{$26$} & \multicolumn{1}{c}{$26$} & \multicolumn{1}{c}{}\\
\end{tabular}
\vspace{0.1cm}
\captionof{table}{Group A=q}
\label{}
\end{minipage}

Note that the predictor is more accurate in both groups. Now we check whether these tables satisfy sufficiency and separation. For sufficiency, we would need that the positive predictive values (PPV) agree, but we have:

\begin{equation}
\frac{a}{a+b} = \frac{11}{13} \neq \frac{21}{25} = \frac{a'}{a'+b'}
\end{equation}

For separation, we would need that the false negative rates (FNR) agree, but we have:

\begin{equation}
\frac{c}{a + c} = \frac{2}{13} \neq \frac{5}{26} = \frac{c'}{a'+ c'}
\end{equation}

Thus, we have increased accuracy and lost both separation and sufficiency. This shows that separation and sufficiency are not incrementally conservative fairness measures. $\square$

Note that if we had increased accuracy in proportion to group size, i.e., if we had shifted two elements instead of one from false negatives to true positives in group $q$, we would have preserved sufficiency and separation. The reason for this is that this increment would have preserved the proportions of the confusion matrices between the two groups. However, this is a very special kind of increment. The case we have discussed above, with increments not proportional to the size of the groups, is easier to realize and presumably more common.

\emph{Proof of proposition \ref{prop: sufficiency}:} Sufficiency for groups $A = p, q$ means, in the case $R=+$ and $Y = +$:

\begin{eqnarray*}
P(Y = + \mid A=p, R=+) &=& P(Y=+ \mid A=q, R = +)\\
\Leftrightarrow \frac{a}{a + b} & = & \frac{a'}{a' + b'},
\end{eqnarray*}

where the choice of $Y = -$ yields an equivalent condition; the same reasoning holds for $R = -$. $\square$

\emph{Proof of proposition \ref{prop: separation}:} Similar to proof of proposition \ref{prop: sufficiency}.

\begin{acks}
I thank Michele Loi, Corinna Herweck, and members of the philosophy of science research colloquium in the Fall of 2020 at the University of Bern for helpful comments on an earlier draft of the paper. This work is
supported by the \grantsponsor{SNF}{National Research Programme ``Digital Transformation'' (NRP 77) of the Swiss National Science Foundation (SNSF)}{} under Grant No.: \grantnum{SNF}{187473}.
\end{acks}

\end{document}